\documentclass[twocolumn,showpacs,preprintnumbers,prb,amsmath,amssymb]{revtex4}

\usepackage{graphicx}
\usepackage{bm}
\usepackage{epsfig}

\newcommand{\Srn}{SrFe$_{2-x}$Ni$_x$As$_2$}
\newcommand{\Src}{SrFe$_{2-x}$Co$_x$As$_2$}
\newcommand{\Ban}{BaFe$_{2-x}$Ni$_x$As$_2$}
\newcommand{\Bac}{BaFe$_{2-x}$Co$_x$As$_2$}
\newcommand{\Sr}{SrFe$_2$As$_2$}
\newcommand{\Ba}{BaFe$_2$As$_2$}
\newcommand{\Niop}{SrFe$_{1.85}$Ni$_{0.15}$As$_2$}

\newcommand{\tc}{$T_c$}

\newcommand{\ie}{{\it i.e.}}

\begin{document}


\title{Evolution of bulk superconductivity in SrFe$_2$As$_2$ with Ni substitution}


\author{S.~R.~Saha}
\author{N.~P.~Butch}
\author{K.~Kirshenbaum}
\author{Johnpierre~Paglione}
\email{paglione@umd.edu}

\address{Center for Nanophysics and Advanced Materials, Department of Physics,
University of Maryland, College Park, MD 20742}

\date{\today}


\begin{abstract}
Single crystals of the Ni-doped FeAs-based superconductor \Srn\ were grown using
a self-flux solution method and characterized via x-ray measurements and low temperature transport, magnetization, and specific heat studies. A doping phase diagram has been
established where the antiferromagnetic order associated with the magnetostructural transition of the parent compound \Sr\ is gradually suppressed with increasing Ni concentration, giving way to bulk-phase superconductivity with a maximum transition temperature of 9.8~K. The superconducting phase exists through a finite range of Ni concentrations centered at $x=0.15$, with full diamagnetic screening observed over a narrow range of $x$ coinciding with a sharpening of the superconducting transition and an absence of magnetic order. An enhancement of bulk superconducting transition temperatures of up to $20\%$ was found to occur upon high-temperature annealing of samples.

\end{abstract}


\maketitle


\section{Introduction}

The appearance of superconductivity in iron-based pnictide compounds
has attracted much attention, providing both a new potential angle
in understanding the physics of high-temperature superconductivity
in other materials such as the copper-oxides, and an entire new
family of superconducting materials of fundamental and technological
interest. Superconductivity with \tc\ = 26~K was first reported in
LaO$_{1-x}$F$_x$FeAs at ambient pressure,\cite{Kamihara} and later
raised to 43~K under applied pressures.\cite{Takahashi} The highest
\tc\ achieved so far in these materials is about 55 K in
SmO$_{1-x}$F$_x$FeAs~(Ref.~\onlinecite{Ren1}) and (Ba,Sr,Ca)FeAsF
(Ref.~\onlinecite{Zhu,Cheng}). Oxygen-free FeAs-based compounds with
the ThCr$_2$Si$_2$-type (122) structure also exhibit
superconductivity but so far at slightly lower temperatures, with a
maximum value of \tc\ $\simeq$37 K induced by chemical substitution
of alkali or transition metal ions \cite{Sasmal,Rotter,Sefat,Leithe}
or by the application of large
pressures.\cite{Torikachvili,CaFe2As2,Alireza,Kumar} A few
stoichiometric FeAs-based 122 compounds, including KFe$_2$As$_2$ and
CsFe$_2$As$_2$ (Refs.~\onlinecite{Rotter2,Sasmal}) show
superconductivity below 4~K at ambient pressures.  Despite the lower
values of \tc, the 122 compounds are an important experimental
platform for understanding Fe-based superconductivity, as it is
possible to synthesize large, high quality single crystals, whereas
it is rather difficult for the 1111 compounds.

It is widely believed that suppression of the magnetic/structural
phase transition in these materials, either by chemical doping or
high pressure, is playing a key role in stabilizing
superconductivity in the
ferropnicitides.\cite{bondangle,Kreyssig,Canfield}  For instance,
superconductivity has been induced by partial substitution of Fe by
other transition metal elements like Co and Ni in both the
1111~(Refs.~\onlinecite{Sefat2, Cao1, Cao2}) and 122
compounds.\cite{Sefat,Leithe} For the 122 phase, superconductivity
with \tc\ as high as 25 K in \Bac\ (Ref.~\onlinecite{Ni,Chu}) and
\Src\ (Ref.~\onlinecite{Leithe}) systems, and 21 K in \Ban\
(Ref.~\onlinecite{Li, Canfield}) has also been observed. Very
recently, Ru, Ir, and Pd substitution for Fe was also shown to
induce superconductivity in polycrystalline \Sr\
samples.\cite{Schnelle,Han,Zhu2} As implied by the enhanced negative
thermoelectric power value in the normal state,\cite{Wu,Li} Co and
Ni substitution appears to donate negative charge carriers that are
thought to lead to superconductivity.  Interestingly, in
\Bac,\cite{Chu,Wang} the maximum \tc\ is found at $x\simeq$0.17,
whereas in \Ban, the maximum \tc\ occurs at approximately $x=$0.10
(Refs.~\onlinecite{Li, Canfield}), suggesting that Ni substitution
may indeed contribute twice as many $d$-electrons to the system as
Co. Regarding this, an important question to ask is whether an
analogous situation exists in a system with different structural
parameters, such as \Sr. While there have been several
studies\cite{Leithe,Chu,Wang} of \Src, no bulk superconductivity has
been reported in \Srn.

To investigate the effects of Ni substitution in an as-yet
unexplored series of the FeAs-based 122 compounds, a study of the
evolution of superconductivity in single-crystalline \Srn\ was
performed. Here we report superconductivity induced by Ni
substitution in the series \Srn\ with maximum \tc\ (onset) of 9.8~K.
By studying a wide range ($x$=0--0.30) of single-crystal samples, we
establish a new member of the 122 series with superconductivity
induced by transition metal substitution for Fe. Contrary to
expectations framed by prior studies of similar compounds, we
observe a relatively low maximal \tc\ value of $\sim 10$~K in this
series, centered at a Ni concentration approximately half that of
the optimal Co concentration in \Src.\cite{Chu,Wang} Below, we
discuss the evolution of electrical transport, magnetic and
thermodynamic quantities as a function of Ni concentration, studying
the characteristics of the doping-induced superconductivity in this
system. We also discuss similarities and differences between this
new superconducting system and other members of the 122 family of
iron-pnictide superconductors.

\begin{figure}[!t]\centering
       \resizebox{8cm}{!}{
              \includegraphics{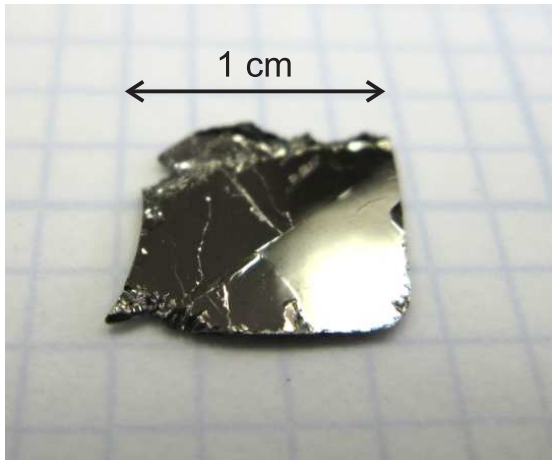}}
              \caption{\label{fig1} Digital image of a typical as-grown single crystal of
              \Srn\ harvested from flux growth. The arrow shows the large platelet dimension,
              indicative of crystals limited by crucible size.}
\end{figure}

\section{Experimental}
Single-crystalline samples of \Srn\ were grown using the FeAs
self-flux method.\cite{Wang} The FeAs and NiAs binary precursors
were first synthesized by solid-state reaction of Fe (5N)/Ni (5N)
powder with As (4N) powders in a quartz tube of partial atmospheric
pressure of Ar. The precursor materials were mixed with elemental
Sr (3N5) in the ratio $4-2x$:$2x$:1, placed in an alumina crucible and
sealed in a quartz tube under partial Ar pressure. The mixture was heated to 1200$^\circ$C, slow-cooled to a lower temperature and then quenched to room temperature.
Figure~\ref{fig1}(a) presents a typical as-grown single crystal
specimen of \Srn\ with $\sim$100~$\mu$m thickness and up to 1~cm width (the size of the crystals was typically found to be limited by the diameter of the crucibles).

Structural properties were characterized
by both powder and single-crystal x-ray diffraction and Rietfeld
refinement (SHELXS-97) to $I4/mmm$ structure. X-ray diffraction
(XRD) was performed at room temperature using a Siemens D5000
diffractometer with Cu-K$\alpha$ radiation, with lattice parameters
refined by a least-squares fit. Chemical analysis was obtained via
wavelength-dispersive x-ray spectroscopy (WDS) and energy-dispersive
x-ray spectroscopy (EDS), showing proper stoichiometry in all
specimens reported herein and no indication of impurity phases.

Resistivity ($\rho$) samples were prepared using gold wire/silver paint contacts made at room temperature, yielding typical contact resistances of $\sim$ 1 $\Omega$.
Resistance measurements were performed using the standard four-probe AC
method, with excitation currents of 1~mA at higher temperatures that were reduced to 0.3~mA at low temperatures to avoid self-heating, all driven at 17 Hz frequency.
Magnetic susceptibility ($\chi$) was measured using a Quantum Design SQUID magnetometer, and specific heat was measured with a Quantum Design cryostat using the thermal relaxation method.

\section{Results and Discussion}

\begin{figure}[!t]\centering
       \resizebox{8.5cm}{!}{
              \includegraphics{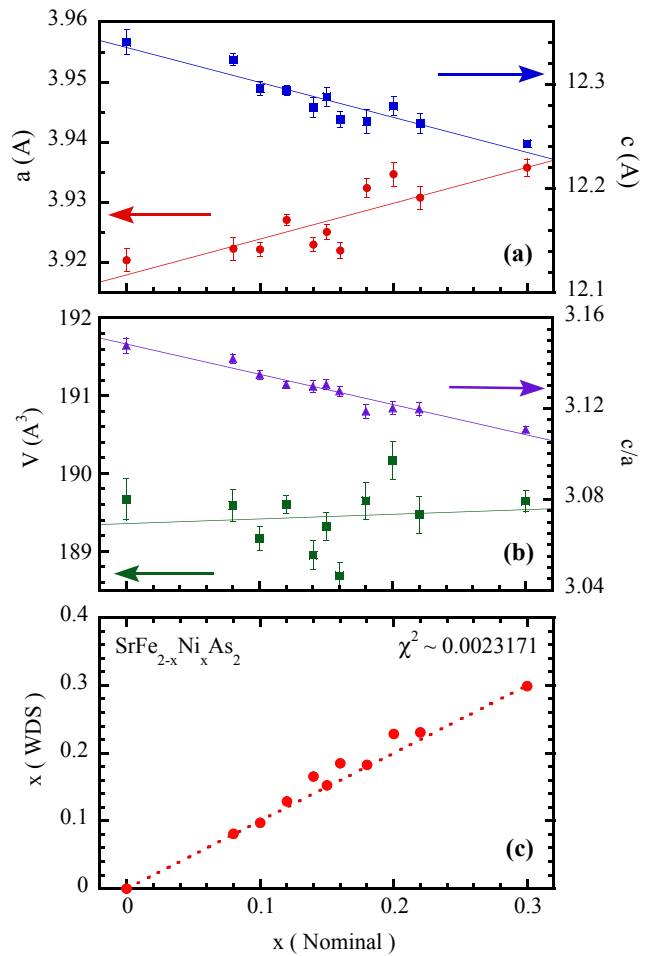}}
              \caption{\label{fig2}(a) Variation of the $a$- and $c$-axis
              lattice constants of \Srn\ with Ni content $x$, as determined
              from Rietfeld refinement of x-ray powder diffraction spectra;
              (b) Corresponding change of tetragonal $c/a$ ratio and unit cell volume $V$;
              (c) Actual Ni concentration of \Srn\ single-crystal samples  as a function of
              nominal concentration $x$, as determined by wavelength dispersive x-ray spectroscopy
              (data points represent average value of 10 scanned points for each concentration,
              the dotted line is a linear fit with a slope of 1).}
\end{figure}

\subsection{Structural and Chemical Characterization}

Fig.~\ref{fig2}(a) presents crystallographic $a$- and $c$-axis
lattice constants determined from refinement fits of x-ray diffraction patterns of powdered samples of \Srn\ as a function of Ni concentration
$x$, along with the resultant tetragonal ratio $c/a$ shown in Fig.~\ref{fig2}(b). With increasing $x$, the $c$-axis lattice constant decreases
and the $a$-axis lattice constant increases, while the $c/a$ ratio
decreases linearly, without any significant change in unit cell
volume to within experimental accuracy. Fig.~\ref{fig2}(c) shows the
actual Ni concentration determination in \Srn\ crystals measured by
WDS analysis, using an average value determined from 10 different
spots on each specimen, plotted as a function of nominal
concentration $x$. Because a linear fit (dotted line) results in a
slope of unity to within scatter, the nominal value of $x$ will be
used hereafter as an adequate representation of the actual
concentration.


\begin{figure}[!t]\centering
       \resizebox{8.5cm}{!}{
              \includegraphics{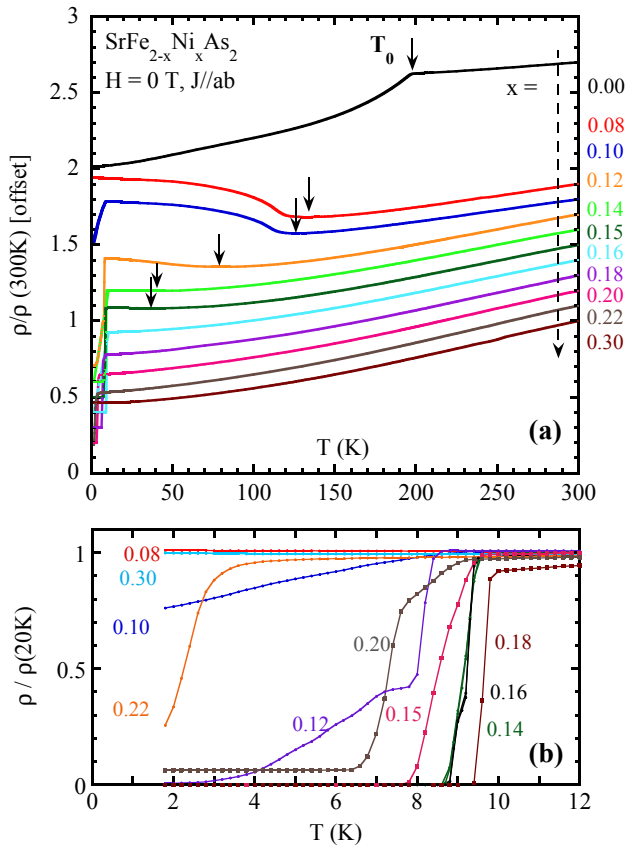}}
              \caption{\label{fig3}(a) Temperature dependence of in-plane electrical
              resistivity of specimens of \Srn, normalized to 300~K and offset for clarity (data sets placed above $x$=0.30 are successively offset
              vertically by 0.1, except for $x$=0 data, which
              are offset by 1.7). The direction of the broken arrow indicates the order of the resistivity
              curves with ascending $x$ as noted to the right. Short arrows indicate the position
              of the magnetic transition $T_0$, defined by the kink in $x=0$ data
              and the minima in $\rho(T)$ data for $0.08 \leq x \leq 0.15$.
              (b) Expanded low-temperature view of resistivity normalized to 20~K for clarity,
              showing the evolution of superconducting transitions with Ni concentration.}
\end{figure}

\subsection{Electrical Resistivity}

\begin{figure}[!t] \centering
  \resizebox{8.5cm}{!}{
  \includegraphics{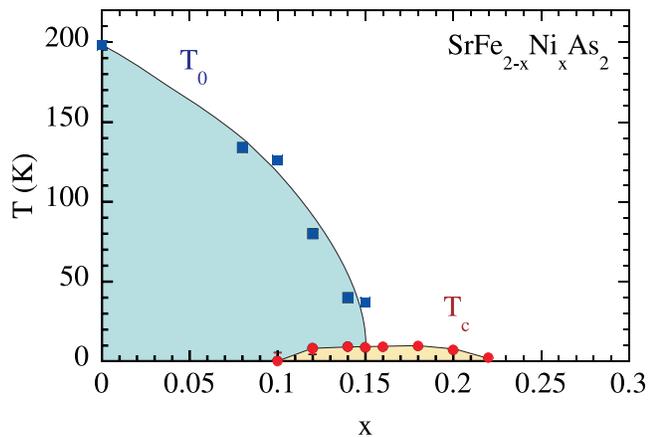}}
  \caption{\label{fig4} Ni substitution phase diagram of \Srn\ obtained from
  electrical resistivity data, showing the suppression of the magnetic/structural phase transition
  $T_0$ (blue squares) with increasing Ni concentration, and the appearance of a superconducting
  transition (red circles) with maximum \tc\ of $\sim $10~K centered around $x\simeq 0.15$.}
\end{figure}

Fig.~\ref{fig3}(a) presents the comparison of the in-plane
resistivity $\rho(T)$ of single crystals of \Srn\ (data are
presented after normalizing to room temperature and offsetting for
clarity). As shown, $\rho(T)$ data for \Sr\ exhibit metallic
behavior, decreasing with temperature from 300~K before exhibiting a
sharp kink at $T_0=198$~K, where a structural phase transition (from
tetragonal to orthorhombic upon cooling) is known to coincide with
the onset of antiferromagnetic (AFM) order.\cite{Yan} With
increasing Ni substitution the anomaly associated with $T_0$ becomes
less distinct and is defined by a smooth minimum in $\rho(T)$, which
shifts to lower temperature as indicated by the position of short arrows in
Fig.~\ref{fig3}(a), finally disappearing for $x >$0.15 where no
minimum is evident.
We define the value of $T_0$ as the position of the kink in $x=0$ data and the position of the minima in $\rho(T)$ data for $0.08 \leq x \leq 0.15$, and present its evolution with Ni concentration in Fig.~\ref{fig4}.

The sharp decrease in $\rho(T)$ associated with $T_0$ in the undoped
material is observed to change character with increased Ni
substitution, as it is shifted to lower temperatures. This switch,
from a drop in $\rho(T)$ to a rise in $\rho(T)$ with decreasing $T$
as $T_0$ is suppressed, has also been observed in other doped 122
materials,\cite{Li,Leithe,Ni,Canfield} and likely arises due to a
shift in the balance between the loss of inelastic scattering due to
the onset of magnetic order and the change in carrier concentration
associated with the transition at $T_0$. Interestingly, the
substitution of Ni for Fe appears to have minimal effect on
inelastic scattering in the paramagnetic state, as indicated by the
identical slope and curvature of all $\rho(T)$ curves above $T_0$ in
Fig.~\ref{fig3}(a). This can be considered as a confirmation of the
dominant role of phonon scattering in determining the temperature
dependence of resistivity.

For $x=0$, $\rho$ continues to decrease below $T_0$ without any
trace of superconductivity down to 1.8 K. (The appearance of
strain-induced superconductivity with $T_c=21$~K has been previously
shown to appear in undoped ($x=0$) samples of \Sr.\cite{Saha}
However here we present $x=0$ data for a sample with all traces of
superconductivity removed by heat treatment.) This is also the case
for $x=$0.08, with no evidence of superconductivity down to 1.8 K.
However, $x=$0.1 begins to show traces of superconductivity as
evidenced by a partial drop in $\rho(T)$ below $\sim$ 10 K as shown
in Fig.~\ref{fig3}(b). For $x=$0.12, there is a sharp drop below
8.4~K that does reach zero resistance at lower temperatures. This
partial transition turns into a full transition for $x \geq $0.14
with higher \tc. In the range of samples studied, the highest \tc\
is obtained for $x=$0.18 with a $\sim$ 9.8 K onset and $\sim$9.6 K
midpoint. For $x \geq $0.2, superconductivity becomes partial again
with incomplete superconducting transitions shown in the $x=0.20$
and $x=0.22$ samples and the complete absence of any superconducting
transition down to 1.8~K for $x=$0.3.

Figure~\ref{fig4} presents the phase diagram representing the variation of $T_0$ and \tc\  (determined as noted above and at the $50\%$ drop of $\rho$, respectively), as a function of Ni content $x$. The superconducting window spans the range $x=$0.1-0.22 (see also
Fig.~\ref{fig7}(d) below for a detailed view) and forms a dome-like
superconducting phase that appears qualitatively similar to other
Co- and Ni-doped 122 compounds.


\begin{figure}[!t] \centering
  \resizebox{8.5cm}{!}{
  \includegraphics{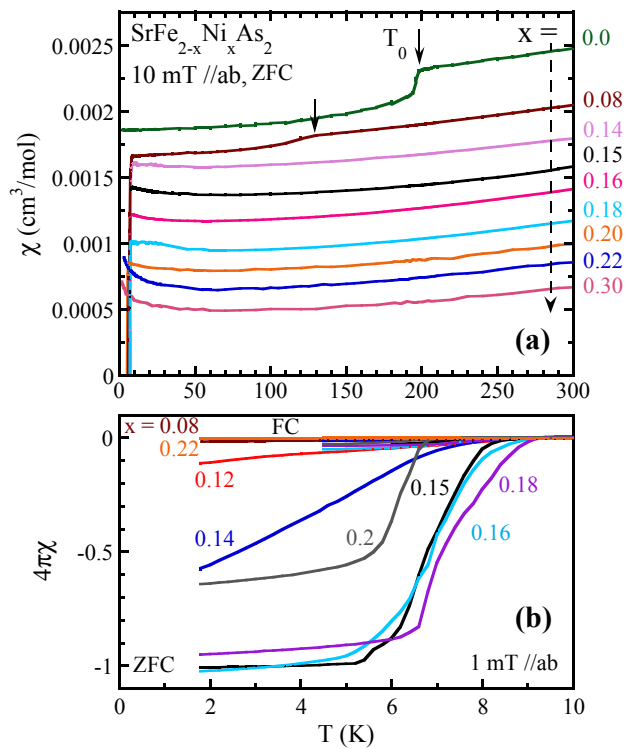}}
  \caption{\label{fig5} Temperature dependence of magnetic susceptibility
  $\chi$ of \Srn, measured with 10~mT field applied parallel to the
  crystallographic basal plane from zero-field-cooled (ZFC) conditions, offset for clarity ($x$=0 data are vertically offset by +0.0015~cm$^3$/mol, with other sets offset successively downward by $\sim 0.0002$).
    (b) Low-temperature zoom of the volume magnetic susceptibility in \Srn\ crystals
  under 1 mT ZFC and field-cooled (FC) conditions after a 24 hour / 700$^\circ$C annealing treatment (see text for details).}
\end{figure}

\subsection{Magnetic Susceptibility}

Figure~\ref{fig5}(a) presents the temperature dependence of magnetic
susceptibility $\chi$ of \Srn\ crystals, measured under
zero-field-cooled (ZFC) conditions by applying 10 mT along the
$ab$-plane. The data are presented with a $y$-axis offset for
clarity purposes ($x$=0 data have been shifted by +0.0015 cm$^3$/mol, and successive data sets for $x>0$ have been staggered downward), however note that absolute values at room temperature for all Ni concentrations are all approximately $\chi(300$~K)$\simeq 0.001$~cm$^3$/mol to within experimental error.
As shown, the overall behavior of low-field susceptibility for
$x=0$ is similar to that reported previously~\cite{Yan} for high-field conditions, showing a modest temperature dependence interrupted by a sharp drop at $T_0$ due to the magnetic/structural transition.
The overall temperature dependence and magnitude of $\chi$ remains more or less
constant with Ni doping, indicating minimal impact of Ni substitution on the paramagnetic susceptibility of \Srn. With increasing Ni concentration, the large step-like feature at $T_0$ is suppressed to lower temperatures and dramatically reduced in
magnitude, as indicated by a small kink at $T_0$ for $x=0.08$ and no
discernible feature for higher $x$. This behavior is comparable to the effect of Co doping in the \Bac\ series,\cite{Ni} which shows a similar trend in magnetization data
taken at 1~T.

Note that the low-field $\chi(T)$ data presented here do
not show any significant increase at low temperatures, indicating both good
sample quality (\ie, minimal magnetic impurity content) and no
indication of strain-induced superconductivity.~\cite{Saha}
A very small upturn in $\chi(T)$ does appears to onset at low temperatures in all Ni-doped samples. Although its magnitude is quite small, the systematic presence of this upturn along with its slight enhancement in higher Ni-doped samples (\ie, $x$=0.22 and 0.30 data sets) suggests the presence of either a small magnetic impurity content or a small local-moment contribution, possibly due to the presence of Ni. Because a Curie-like tail was reported in \Sr\ samples even at high (5~T) fields, albeit with a much more pronounced increase at low temperatures,\cite{Yan} impurity contributions are less likely. In any case, more work is required to discern the origin of this feature.

Shown in Fig.~\ref{fig5}(b) are the low temperature susceptibility
data for \Srn\ samples measured with a smaller applied field of 1 mT along
the $ab$-plane under both ZFC and field-cooled (FC) conditions,
plotted as the volume susceptibility 4$\pi \chi$ to compare the
level of diamagnetic screening due to superconductivity.  As shown,
the superconducting volume fraction, as estimated by the fraction of
full diamagnetic screening ($4\pi \chi=-1$), varies with Ni
concentration, being absent for $x<0.12$, partial for $x=$0.12,
0.14, and 0.20, and complete for $x=$0.15, 16 and 0.18. This
suggests that there is indeed a full superconducting volume fraction
observed for a range of Ni concentrations with maximized $T_c$
values, but also that partial volume fractions are evident at the
fringes of the superconducting dome. For instance, note that a drop
in $\chi(T)$ is visible below 7 K in the $x=0.08$ data shown in
Fig.~\ref{fig5}(a), but also that the volume fraction associated
with this diamagnetic screening is very small, being less than
$\sim1\%$ as evident from Fig.~\ref{fig5}(b). Likewise, data for
$x=0.12$ show a somewhat larger response but still remain at much
less than 100\%. This is quantified in Fig.~\ref{fig7} in comparison
to other quantities of interest, as discussed below.


\begin{figure}[!t] \centering
  \resizebox{8.5cm}{!}{
  \includegraphics{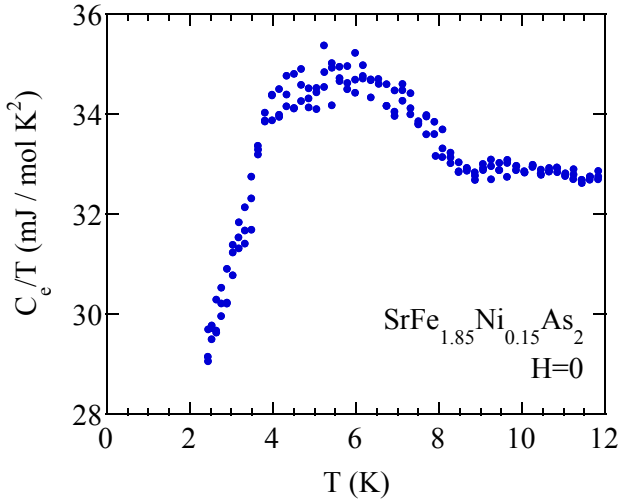}}
  \caption{\label{fig6} Temperature dependence of the electronic specific heat $C_e$ in \Niop, obtained
  by standard fitting and subtraction of a phonon contribution to the total specific heat (see text).
  Data show a small but distinct transition consistent with superconductivity below $T_c=8.5$~K determined
  from magnetic susceptibility measurements.}
\end{figure}

\subsection{Specific Heat}

To verify the bulk thermodynamic nature of the superconducting
transition in \Srn, we performed specific heat measurements on an
annealed sample with optimal Ni concentration of $x=0.15$. The
electronic specific heat $C_e$ was determined by subtracting the
phonon contribution from the total specific heat at zero magnetic
field. Fitting of the $x=$0.15 data to the standard form $C_p = \gamma T + \beta T^3$ for the total specific heat through the range 75~K$^2$ $\leq$ $T^2$ $\leq$ 290~K$^2$ yields an electronic contribution $\gamma = $32~mJ/mol~K$^2$ and a phononic contribution $\beta =$ 0.76~mJ/mol~K$^4$, the latter value corresponding to a Debye temperature of $\Theta_D =$ 234~K. For $x=$0 (not shown), $C_e/T$ is almost independent
of $T$ at low temperatures down to 2 K to within the experimental
accuracy, exhibiting comparable fit parameters to those above and
thus verifying that no significant change of the phonon spectrum is
imparted by Ni substitution.

Fig.~\ref{fig6} presents the low temperature portion of $C_e/T$ for
$x=$0.15, highlighting the onset of a weak anomaly below 8.5 K that
is consistent with the value of \tc\ deduced from $\chi(T)$
measurements. Although the peak in $C_e/T$ is too poorly defined to
fit with an equal entropy construction, a rough quantitative
characterization provides an estimated value of $\Delta C/\gamma T_c
\simeq 0.12$. This is much smaller than the BCS expectation of 1.52
for a superconducting transition, but is not surprising considering
the similar trend found in the literature. Although a sizeable
specific heat anomaly has been observed at the superconducting
transition of some Fe-based superconductors, including values near
the BCS expectation in both Ba$_{0.6}$K$_{0.4}$Fe$_2$As$_2$
(Ref.~\onlinecite{Mu2}) and LaFePO (Ref.~\onlinecite{baumbach}),  it
is intriguing that many members of the FeAs-based family --
including both Co-doped \Ba\ (Ref.~\onlinecite{Chu}) and
CaFe$_2$As$_2$ (Ref.~\onlinecite{NKumar}), and F-doped LaFeAsO
(Ref.~\onlinecite{Mu}) and SmFeAsO (Ref.~\onlinecite{Ding}) --
exhibit rather weak signatures of superconductivity in specific heat
measurements, despite indications of bulk diamagnetic screening from
magnetization measurements.

Likewise, the anomaly in the specific heat of the $x=$0.15 sample of
\Srn\ shown in Fig.~\ref{fig6} is surprisingly small. The small peak
observed in several of these materials would normally seem to
reflect a need to improve sample quality via improved growth
techniques or annealing treatment, as was indeed shown for the case
of LaFePO.\cite{baumbach}  However, measurements of $C_e/T$ in an
as-grown, unannealed $x=0.15$ sample of \Srn\ (not shown) also
present a weak feature at \tc\ quantitatively comparable in magnitude to that
discussed above for the annealed sample. In light of the enhancement
of \tc\ invoked by annealing discussed below, this suggests that, at
least for the case of \Srn, improvements in the superconducting
properties do not lead to enhanced values of $\Delta C/\gamma T_c$
as would be expected for improved sample quality. More important,
the small values of $\Delta C/\gamma T_c$ observed in many
FeAs-based materials are difficult to reconcile with consistent
observations of bulk diamagnetic screening, including those for many
\Srn\ samples in this study of widely varying size and shape.
Overall, this suggests that a lack of sample quality may not always
be responsible for poor thermodynamic signatures of
superconductivity in these materials, and that alternative
explanations should not yet be ruled out. For instance, the small
size of $\Delta C/\gamma T_c$ in \Srn\ and the large residual
density of states may imply that superconductivity gaps only a small
part of the Fermi surface.

\subsection{Doping Evolution}

\begin{figure}[!t] \centering
  \resizebox{8.5cm}{!}{
  \includegraphics{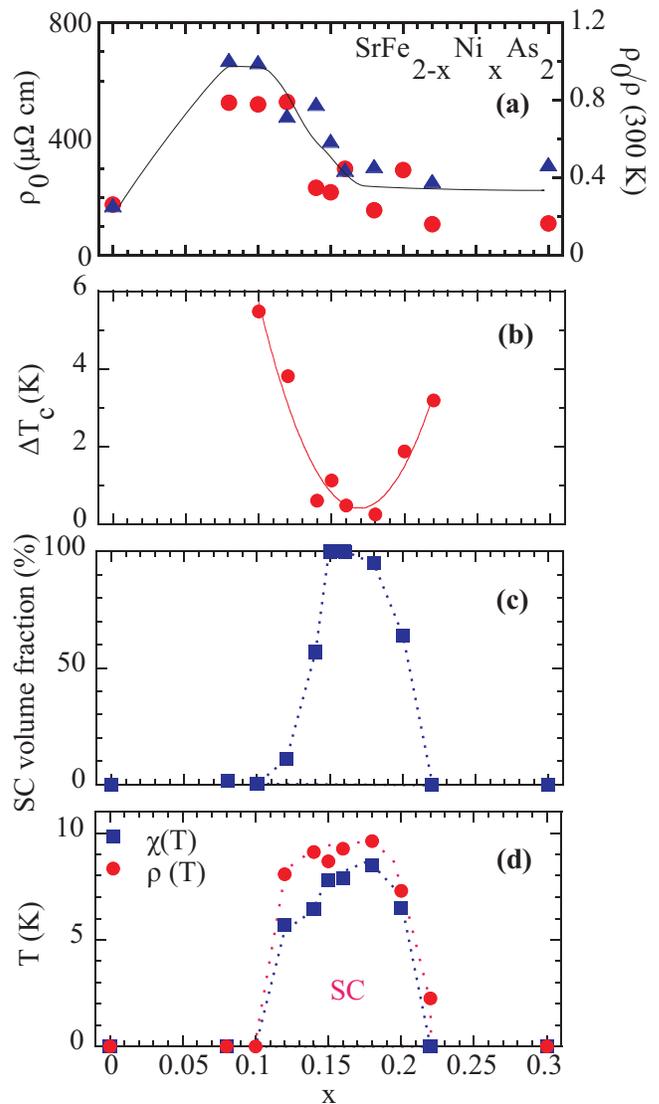}}
  \caption{\label{fig7} Evolution of normal and superconducting state parameters in \Srn\ with Ni concentration $x$:
  (a) absolute (circles-left scale) and normalized (triangles-right scale) residual resistivity;
  (b) variation of the width in temperature of the resistive superconducting transition $\Delta$\tc\ .
  (c) superconducting volume fraction determined from the magnetic susceptibility data.
  (d) expanded view of the superconducting phase determined by transition temperatures
  defined by $50\%$ resistivity drop (circles) and $10\%$ value of total diamagnetic screening (blue squares).
  All lines are guides.}
\end{figure}

It is instructive to compare the evolution of the superconducting
state parameters in more detail as a function of Ni concentration.
In Fig.~\ref{fig7}, we compare measures of the width of the
superconducting transition $\Delta T_c$ as defined by the difference
of \tc\ at $90\%$ and $10\%$ drop of resistivity from its normal
state value, the estimated superconducting volume fraction
determined from the level of diamagnetic screening, and the
evolution of \tc\ itself as determined by transitions in both
resistivity and susceptibility. These parameters are plotted
alongside the values of residual resistivity $\rho_0(x)$ (determined
by linear extrapolations from above $T_c$) to compare the evolution
of superconductivity with residual transport scattering behavior,
also used as a measure of where magnetic order is suppressed.

The evolution of $\rho_0$ with Ni doping is plotted in
Fig.~\ref{fig7}(a), including both the absolute value of $\rho_0$
(left $y$-axis) and that normalized by $\rho$(300~K) (right $y$-axis) to
remove uncertainty in geometric factors. As a function of
$x$, both absolute and normalized values of $\rho_0(x)$ follow a
similar pattern, suggesting that geometric factor errors are not
large. As shown in Fig.~\ref{fig7}(a), an increase in resistivity
occurs with increasing Ni concentration between $x=0$ and 0.08
before showing an approximate plateau up to $x=0.12$, reflecting the
change in transport scattering associated with magnetic order at low
concentrations. Above $x=0.12$, $\rho_0$ shows a rapid decrease with
increasing $x$ before again leveling off at higher Ni concentration,
coincident with the complete suppression of magnetic order near
$x=0.15$ and the onset of superconductivity. This trend follows what
can be inferred from the $\rho(T)$ data found in Fig.~\ref{fig3}(a),
with an enhancement of $\rho_0$ found only in the regime
($0<x<0.12$) where inelastic scattering is greatly enhanced by the
presence of magnetic order, resulting in an increase in $\rho(T)$
below $T_0$.

Interestingly, aside from this enhancement, the
impurity scattering level (as measured by the value of $\rho_0(x)$)
does not show any significant change with Ni concentration, with
values of $\rho_0$ in high Ni content samples approaching that of
$\rho(x=0)$. In a minimal model where residual resistivity is
dominated by impurity/disorder scattering, this trend would suggest
that Ni substitution for Fe introduces minimal disorder into the
system, even up to $x=0.30$ levels. However, it is likely that a
more unconventional mechanism (such as magnetic fluctuation
scattering) may dominate the value of $\rho_0$ in this system,
thereby masking the underlying (small) increase in residual
scattering due to Fe site disorder.

A detailed plot of \tc\ vs. $x$ is presented in Fig.~\ref{fig7}(d),
showing good agreement between \tc\ values determined by transitions
in $\rho(T)$ and $\chi(T)$. As is evident from the comparison of
$\rho_0(x)$ to $T_c(x)$ in Fig.~\ref{fig7}, the rather abrupt decrease
in residual scattering occurs very close to the appearance of bulk
superconductivity in \Srn.  The Ni concentration of $x=0.14$ is
where $\rho_0$ drops to its low value and a sizeable volume fraction
of superconductivity first appears, as shown in Fig.~\ref{fig7}(b).
Both the width $\Delta T_c$ of the transition and the
superconducting volume fraction change dramatically in this
concentration range. As shown, there is an
interesting inverse correlation between $\Delta T_c$ and this volume
fraction within the range of superconducting samples, illustrating
that the sharpest superconducting transitions are associated with
bulk superconductivity, while the broader transitions are associated
with only partial volume superconductivity.

With the current set of measurements, it is hard to distinguish
whether there is an inhomogeneous distribution of Ni content in the
samples close to this boundary causing the  partial superconducting
transitions, or whether the narrow range of bulk superconductivity
is truly an intrinsic property. However, several factors suggest
that inhomogeneity in Ni concentration should not be significant.
First, x-ray diffraction and chemical analysis data presented in
Fig.~\ref{fig2} suggest that Ni substitution is occurring smoothly
and continuously in this series, with no observable deviations at
the edges of the superconducting dome. Second, both resistivity and
susceptibility data presented above also progress smoothly as a
function of $x$, again indicating no major levels of phase
separation.  Finally, note that all \Srn\ crystalline samples used
in this study have been annealed at high temperatures to further
reduce smaller inhomogeneity levels, as discussed below.

\subsection{Annealing Effect on Superconductivity}

\begin{figure}[!t] \centering
  \resizebox{8cm}{!}{
  \includegraphics{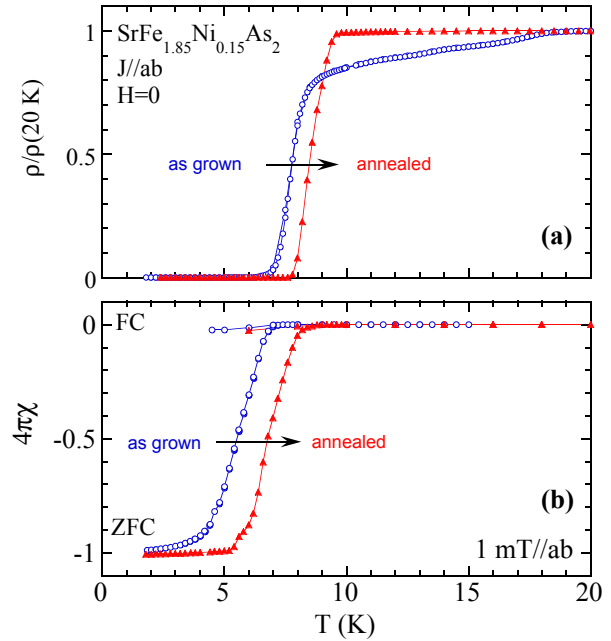}}
  \caption{\label{fig8}Effect of high-temperature annealing on an optimally doped $x=0.15$ sample of \Srn,   demonstrating typical results from before and after a 24 hour, 700$^\circ$C heat treatment performed   on a sample sealed in a quartz tube with a pure argon environment. (a) Resistivity data of a $x=0.15$ sample measured before (blue circles) and after (red triangles) heat treatment. (b) Volume magnetic susceptibility of a $x=0.15$ sample at low fields measured before (blue circles) and after (red triangles) annealing. Arrows emphasize enhancement of \tc\ by annealing, with good agreement in \tc\ values for both cases.}
\end{figure}

One method of investigating the effect of crystalline quality is by high-temperature heat treatment. Interestingly, we found that annealing single crystals of \Srn\ in such a way produces a rather dramatic enhancement in the value of \tc: holding samples at 700$^\circ$C for 24 hours in an Ar atmosphere was found to increase \tc\ by up to $\sim 1$~K. As shown in Fig.~\ref{fig8}, the effect of annealing on the superconducting transition in \Niop\ crystals is evident in both $\rho(T)$ and $\chi(T)$, indicating that this enhancement is reflected in the full diamagnetic screening and is therefore a bulk phenomenon. Such an enhancement of \tc\ could be an indication of improved crystallinity due to release of residual strain, and/or improved microscopic chemical homogeneity of Ni content inside the specimens, thereby optimizing the stability of superconductivity. A similar annealing effect was reported in $Ln$FeOP (Ln=La, Pr, Nd) single crystals, where a heat treatment in flowing oxygen was also found to improve superconducting properties.~\cite{baumbach}

It is further noteworthy to report that as-grown crystals of \Srn\ for $x<0.16$ show what looks to be a partial superconducting transition near 20~K  that is completely removed by heat treatment, as demonstrated in Fig.~\ref{fig8}(a) for $x=0.15$. Although it is tempting to posit that 20~K is a possible value for optimal \tc\ in this series of Ni-substituted compounds, note that aside from the enhancement of \tc\ as mentioned above, the removal of this feature is the only change observed in measured quantities imposed by annealing: neither the resistivity nor the magnetic susceptibility in the normal state show any change after annealing. Furthermore, susceptibility does not show any indication of diamagnetic screening above bulk \tc\ values in the as-grown samples. Because the 20~K kink is removed with heat treatment, and, moreover, is always found to be positioned near the same temperature, we believe this feature may be connected to the strain-induced superconductivity found in undoped \Sr.~\cite{Saha} However, note that whereas only a mild 5-minute heat treatment of 300$^\circ$C removes the partial volume superconductivity in \Sr, a substantially higher-temperature 700$^\circ$C treatment is required to remove this feature in \Srn. If the two phenomena are related, it is possible that internal strain is stabilized by the chemical inhomogeneity associated with transition metal substitution in \Srn\ thus requiring higher temperatures to be removed. More systematic studies of the effect of annealing on \Srn\ are under way to investigate this relationship.


\subsection{Comparison to Other FeAs-Based Systems}

Superconductivity appears in \Srn\ through the range $x=0.1-0.22$, tracing out a dome-like \tc\ curve qualitatively similar to other transition metal-substituted FeAs-based superconducting systems. Naively, in a rigid band model it would be expected that each Ni$^{2+}$ dopant introduces two extra itinerant 3$d$ electrons while each Co$^{2+}$ dopant adds only one. In \Srn, the superconducting phase is centered about an ``optimal'' Ni concentration of $x\simeq 0.15$ that corresponds to 7.5\% Ni substitution for Fe, which is indeed approximately half of the median concentration of Co ($\sim 0.25$-$0.30$) which induces superconductivity in \Src\ through the range $0.15<x<0.40$.\cite{Leithe} This is comparable to the case of \Ban\ and \Bac, where the superconducting phases are centered on $x\simeq0.10$ and $x\simeq0.17$, respectively,\cite{Li,Chu,Canfield} also roughly following the $d$-electron counting trend. What is intriguing, however, is that the absolute percentage of Fe substitution required to induce superconductivity in Sr- and Ba-based 122 systems by the same dopant atom appears to be different. In \Srn, the optimal Ni concentration of $\sim 7.5\%$ is at least $\sim 1.5$ times the optimal Ni concentration in both \Ban, where $x\simeq0.10$ corresponds to 5\% Fe substitution,\cite{Li,Canfield} and the related 1111 compound LaFe$_{1-x}$Ni$_{x}$AsO, where $x\simeq0.04$ corresponds to 4\% Fe substitution.\cite{Cao2}

\begin{figure}[!t] \centering
  \resizebox{8.5cm}{!}{
  \includegraphics{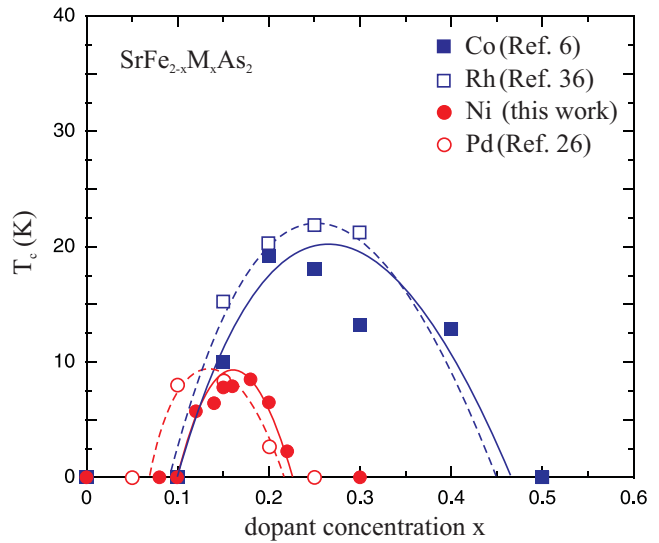}}
  \caption{\label{fig9} Comparison of the evolution of superconductivity as a function
  of Ni substitution in \Srn\ as compared to that previously observed in other transition
  metal substitution series, with M = Co, Rh, and Pd (Refs.~\onlinecite{Leithe,Han2,Zhu2}, respectively).
  Solid symbols denote \tc\ values for 3$d$-electron substituents Co (blue square) and Ni (red circle),
  and open symbols denote those of 4$d$ substituents Rh (blue square) and Pd (red circle).}
\end{figure}

Interestingly, the {\it onset} of superconductivity in Co- and Ni-doped \Sr\ appears to occur near the same substitution concentration of $x\simeq 0.1$, but with \tc\ much suppressed in the \Srn\ system relative to that of \Src. This trend also appears to hold to some degree in the doped \Ba\ system, where the onset concentration for \Ban\ is approximately the same as that of \Bac, while its maximum \tc\ value is somewhat reduced.\cite{Canfield} However the comparison between Ba- and Sr-based 122 materials may not be so straightforward owing to the different alkali earth ions involved. Instead, it is simpler to directly compare the effect of substituting different 3$d$ and 4$d$ metal substitutions in the same Sr-based parent material \Sr. Shown in Fig.~\ref{fig9} is a comparison of the evolution of the superconducting phase in \Srn\ as compared to that of three other characteristic substitution series: Co-doping,\cite{Leithe} Rh-doping,\cite{Han2} and Pd-doping,\cite{Zhu2} providing a complete comparison of the effects of $d$-electron doping with 3$d$- vs. 4$d$-electrons.
Notably, the trend noted above is strikingly similar in the Rh/Pd comparison, which also point to the same onset concentration of $x\simeq 0.1$ and a maximum \tc\ in the Pd-doped system that is also greatly reduced as compared to the Rh-doped system, reaching only $\sim 9$~K (Ref.~\onlinecite{Zhu2}) as compared to $\sim 22$~K (Ref.~\onlinecite{Han2}).

The comparable trends in these two sets of systems raise questions
as to the nature of 1) the similar onset concentration in all
compounds, and 2) the inhibited \tc\ values in the
two-electron-doped systems (\ie, Ni and Pd) as compared to the
one-electron-doped systems (\ie, Co and Rh). One possible
explanation lies in the differences in structural parameters as a
function of doping. In \Srn, the lattice constants increase along
$a$-axis and decrease along the $c$-axis as a function of $x$,
similar to the behavior for substituting Co, Pd, and Ru in
\Sr.\cite{Leithe,Zhu2,Han2} Also, the variation of $c/a$ ratio with
$x$ in \Srn\ is close to that in \Src,\cite{Leithe} although the
maximum value of \tc\ is higher in the latter. On the other hand,
the variation of $c/a$ ratio with $x$ in \Srn\ is different from
that found in SrFe$_{2-x}$Pd$_x$As$_2$,\cite{Zhu2} while the maximum
value of \tc\ is similar in these nominally isoelectronic systems.
In other words, there is no obvious correlation between \tc\ and
$c/a$ ratio, at least in the \Sr\ derived superconductors, that
could explain these phenomena. However, note that the shape of the distorted tetrahedral environment of Fe, likely an important structural parameter, may not have such a simple correlation with lattice parameters and may depend on how the As $z$-coordinate
changes with doping.

It is also important to consider the role of magnetism in stabilizing
superconductivity in the FeAs-based materials. The related and
widely perceived picture is that doping electrons or holes into the
parent phase gradually suppresses magnetic order, with pairing
arising through the inter-pocket scattering of electrons via
exchange of AF spin fluctuations.\cite{Mazin,Kuroki,Wang4,QHan}
Alternatively, magnetic order and superconductivity may compete to
gap similar parts of the Fermi surface, with superconductivity only
appearing when magnetic order is suppressed.  Either way, there is
no doubt that superconductivity is strongly coupled, directly or
indirectly, to the suppression of magnetic order in the FeAs-based
122 systems. As presented previously in Fig.~\ref{fig4},
superconductivity in \Srn\ indeed appears through a range of Ni
concentrations close to where magnetism is suppressed, similar to
several other
systems.\cite{Leithe,Sefat,Ni,Chu,Canfield,Schnelle,Han,Zhu2}  In
\Srn, the critical concentration appears to sit close to the optimal
doping concentration of $x\simeq 0.15$; it is of obvious interest to
determine this value to a more precise degree, along with that for
the other transition metal-substituted series discussed above. This
will require better methods of determining the magnetic transition
temperature $T_0$, as is possible via neutron scattering
experiments.

Interestingly, recent evidence of coexistent magnetic and
superconducting phases on the ``underdoped'' side of the \tc\ dome
in \Bac\ point to a competitive coexistence of these
phases.\cite{Pratt} The onset of $T_c$ in \Srn\ appears to be rather
abrupt, at least more so than the smooth onset observed in
\Bac.\cite{Ni,Chu} This may be due to a number of factors or
differences between these systems, however it is tempting to posit
that superconductivity and magnetism are more antagonistic in this
system than in its Co-doped counterpart. In any case, it will be
important to compare and contrast the detailed nature of these phase
diagrams in order to gain a better understanding of nature of the
interplay of magnetism and superconductivity.

Finally, it is interesting to note that superconductivity appears to
occur over much narrower doping ranges in both Ni- and
Pd-substituted 122 systems, with lower maximum \tc\ values in Ni
(Pd) substituted materials as compared to Co (Rh) substitution.
Together, these contrasts may indicate that the doping ranges that
induce superconductivity may not only be simply shifted by effective
$d$-electron doping level, but may also involve an inherent
suppression of \tc\ that increases with deviations from the
presumably ideal Fe $d$-shell configuration, possibly due to details
of a chemical nature. Such a picture is indeed consistent with the
recent study of Cu-doping in \Ba,\cite{Canfield} where Cu is assumed
to supply three additional $d$-electrons and thereby deviate
strongly from the Fe $d$-shell configuration. Conversely, studies of
Ru-doped \Sr,\cite{Schnelle} involving nominally isovalent Fe
substitution, support the scenario where superconductivity is most
favored by transition metal substitutions that minimally disrupt the
Fe electronic environment. Of course, one must note that
superconductivity is also known to be present in the fully
Ni-substituted end-member SrNi$_2$As$_2$ (a low-temperature
superconductor with $T_c =0.7$~K),\cite{Bauer} although its
relationship to the superconductivity in \Srn\ is unclear. In any
case, this puzzling point certainly warrants further investigation,
for instance via careful inspections of the phase diagrams arising
in single crystals using other transition metal substituents, and
the role of crystalline quality and disorder in suppressing
superconductivity.

\section{Summary}
In summary, single crystals of the Ni-substituted series \Srn\ were
successfully synthesized,  allowing a determination of the phase
diagram across which magnetostructural order is suppressed and
superconductivity arises over a finite window. Upon suppression of
magnetism, a phase of bulk superconductivity centered near an
optimal concentration of $x\simeq 0.15$ is established with \tc\
values reaching as high as $\sim$ 9.8 K. Interestingly, annealing
treatments of as-grown crystals result in a significant
enhancement of up to $20\%$ in superconducting transition temperatures across this
range. In comparison to
its Co-doped counterpart, the observed superconducting phase in
Ni-doped \Srn\ is intriguingly narrow and strongly suppressed, but
it shows similarities to other transition metal-doped systems
undergoing equivalent $d$-electron substitution, suggesting that
similar underlying physics is at play in stabilizing
superconductivity in several FeAs-based materials.

\vskip 0.5cm
\begin{center}
ACKNOWLEDGEMENTS
\end{center}
The authors acknowledge P.~Y.~Zavalij and B.~W.~Eichhorn for
experimental assistance, and P. Bach, K. Jin, X. Zhang, and R.~L.~Greene for useful discussions. N.~P.~B. acknowledges support from a CNAM Glover fellowship.


\end{document}